\documentclass[letterpaper, 10 pt, conference]{ieeeconf}  

\IEEEoverridecommandlockouts                              

\usepackage{amsmath} 
\usepackage{amsfonts}
\usepackage{graphicx}
\usepackage{color} 
\usepackage{booktabs} 
\usepackage{comment}

\makeatletter
\let\MYcaption\@makecaption
\makeatother
\makeatletter
\let\@makecaption\MYcaption
\makeatother

\newcommand{\R}[1]{\mathrm{#1}}         
\newcommand{\B}[1]{\if#1\relax\bm{#1}\else\mathbf{#1}\fi} 
\newcommand{\BB}[1]{\mathbb{#1}}        
\newcommand{\C}[1]{\mathcal{#1}}        
\newcommand{\abs}[1]{\left\lvert #1 \right\rvert}
\newcommand{\norm}[1]{\left\lVert #1 \right\rVert}
\newcommand{\overbar}[1]{\mkern 1.5mu\overline{\mkern-1.5mu#1\mkern-1.5mu}\mkern 1.5mu} 

\newcommand{\de}{\R{d}}     

\newcommand{\targetdistribution}{\rho^\mathrm{T}}                
\newcommand{\singletarget}{T}                           
\newcommand{\massoftargets}{M^\mathrm{T}}                        
\newcommand{\herderdistribution}{\rho^\mathrm{H}}                
\newcommand{\herders}{\B{H}}                                
\newcommand{\singleherder}{H}                                       
\newcommand{\numberofherders}{N^\mathrm{H}}                      
\newcommand{\numberoftargets}{N^\mathrm{T}}                      
\newcommand{\massofherders}{M^\mathrm{H}}                        
\newcommand{\velocityfield}{V}                          
\newcommand{\diffusioncoefficient}{D}                   
\newcommand{\controlaction}{u}                          
\newcommand{\controlactions}{\B{u}}                     
\newcommand{\interactionkernel}{f}                      
\newcommand{\kernelinteractionlength}{L}                
\newcommand{\desiredtargetdistribution}{\overbar{\rho}^\mathrm{T}}   
\newcommand{\reward}[1][]{\if\relax\detokenize{#1}\relax r\else r^{({#1})}\fi}
\newcommand{\domain}{\mathcal{S}}                            
\newcommand{\herderssettlingtime}{T_\mathrm{a,2\%}^\mathrm{H}}            
\newcommand{\targetssettlingtime}{T_\mathrm{a,2\%}^\mathrm{T}}            
\newcommand{\targetdistributionerror}{e^\mathrm{T}}              
\newcommand{\herdermaxvelocity}{v_{\text{max}}}         

\newcommand{\interactionstrengthgain}{K}                
\newcommand{\targetdistributionss}{\rho^{\mathrm{T},\text{ss}}}  
\newcommand{\timedomain}{\BB{R}_+}                      

\newcommand{\timehorizon}{T_\mathrm{h}}                          

\newcommand{\Lnorm}{{\mathcal{L}^2}}

\usepackage{cite}
\makeatletter
\let\NAT@parse\undefined
\makeatother
\usepackage{hyperref}
\hypersetup{%
    colorlinks = true,
    linkcolor = blue,
    anchorcolor = red,
    citecolor = red,
    filecolor = red,
    urlcolor = blue}
\newcommand{\textcite}[1]{\cite{#1}}    

\newcounter{assumption}
\renewcommand{\theassumption}{\arabic{assumption}}

\newcounter{problem}

\newenvironment{problem}{%
  \refstepcounter{problem}%
  \par\noindent\textbf{Control Problem}\itshape\ignorespaces
}{\par\ignorespacesafterend}

\title{\LARGE \bf
Sparse shepherding control of large-scale multi-agent systems via Reinforcement Learning
}

\author{Luigi Catello$^{1}$, Italo Napolitano$^1$, Davide Salzano$^2$, Mario di Bernardo$^{1,2 *}$
\thanks{The authors acknowledge support from the Italian Ministry of University and Research (MUR) under project PRIN 2022 ``Machine-learning based control of complex multi-agent systems for search and rescue operations in natural disasters (MENTOR).''}
\thanks{$^{1}$ Luigi Catello, Italo Napolitano and Mario di Bernardo are with the Modeling and Engineering Risk and Complexity Department, Scuola Superiore Meridionale, via Mezzocannone 4, 80138, Naples, Italy (email: l.catello@ssmeridionale.it, i.napolitano@ssmeridionale.it)}%
\thanks{$^{2}$Davide Salzano and Mario di Bernardo are with the Department of Information Technology and Electrical Engineering, University of Naples Federico II, Naples, Italy (email: davide.salzano@unina.it, mario.dibernardo@unina.it).}%
\thanks{* Corresponding author}
}

\begin{document}

\maketitle
\thispagestyle{empty}
\pagestyle{empty}

\begin{abstract}
We propose a Reinforcement Learning framework for sparse indirect control of large-scale multi-agent systems, where few controlled agents shape the collective behavior of many uncontrolled agents. The approach addresses this multi-scale challenge by coupling ODEs (modeling controlled agents) with a PDE (describing the uncontrolled population density), capturing how microscopic control achieves macroscopic objectives. Our method combines model-free Reinforcement Learning with adaptive interaction strength compensation to overcome sparse actuation limitations. Numerical validation demonstrates effective density control, with the system achieving target distributions while maintaining robustness to disturbances and measurement noise, confirming that learning-based sparse control can replace computationally expensive online optimization.
\end{abstract}

\section{Introduction}
Complex multi-agent systems, ranging from biological swarms to robotic collectives, exhibit emergent behaviors arising from local interactions among individual agents \cite{dorigo2020reflections, dsouza2023controlling}.
While traditional control strategies directly manipulate subsets of agents or their interaction networks \cite{dsouza2023controlling}, many real-world applications, such as search and rescue \cite{lien2009interactive}, crowd evacuation \cite{albi2013modeling}, and environmental management \cite{zahugi2013oil}, require \emph{indirect control}. In this setting, a population of leaders is controlled to influence another population of targets through their interactions \cite{licitra2019single, bernardi2021macroscopic}.
The \emph{shepherding control problem}, in which a group of herders drives a group of targets toward desired spatial configurations, exemplifies this form of indirect control. It has been extensively studied in small-scale systems through explicit modeling of individual agent dynamics \cite{long2020comprehensive, strombom2014solving, napolitano2025hierarchical}.

As the number of agents increases, agent-level descriptions become intractable, motivating the use of continuum models that describe the evolution of population densities through suitable partial differential equations (PDEs) \cite{lama2025nonreciprocal, maffettone2025leaderfollower, albi2025micromacro}. A particularly relevant scenario is the \emph{sparse control} setting, where a small number of herders must steer a large population of targets. This naturally leads to hybrid, or \textit{micro–macro}, formulations that retain agent-based descriptions for the herders while modeling the target population at the macroscopic level \cite{albi2013modeling}.

This work addresses the sparse indirect control challenge by designing optimal control laws that indirectly shape target density distributions. Although recent sparse optimal mean-field control formulations have shown promising results \cite{ascione2023meanfield, albi2016invisible, albi2021optimized}, their computational cost often hinders real-time applicability. 
Alternatively, approaches based on Markov-chain descriptions on discretized domains \cite{elamvazhuthi2021controllability, kakish2022using} can reduce computational burden but are generally restricted to single-herder scenarios with coarse spatial discretizations. To overcome these limitations, we propose a learning-based framework that coordinates multiple herders in continuous domains without requiring expensive online optimization. The effectiveness of the proposed approach is demonstrated through extensive numerical validation.
We wish to emphasize that, while shepherding provides a canonical example of indirect control, the same problem structure--where few controlled agents influence many uncontrolled ones--arises across diverse domains, such as cell organization \cite{yu2021living} and traffic control \cite{stern2018dissipation}. 

\begin{figure*}[t]
    \centering
    \includegraphics[width=0.95\linewidth]{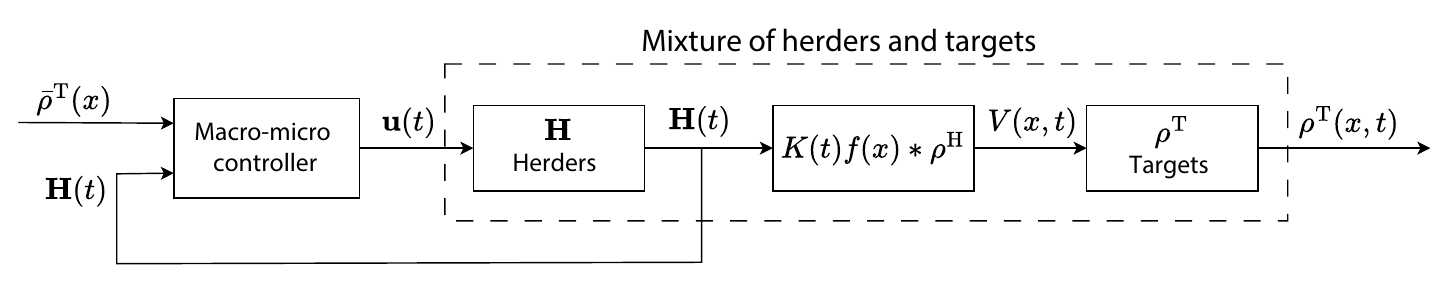}
    \caption{Proposed control architecture. The \emph{macro-micro} controller computes the control actions $\controlactions (t)$ knowing the desired target distribution $\desiredtargetdistribution(x)$ and sensing the current herders' position $\herders (t)$. The herders influence the target density $\targetdistribution (x,t)$ through the velocity field $\velocityfield(x,t)$, thereby achieving indirect control of the target population.}
    \label{fig:control_arch}
    \vspace{-0.35cm}
\end{figure*}

\section{Modeling and problem statement}

We consider the sparse indirect control problem where $\numberofherders$ controlled agents (herders) must shape the spatial distribution of $\numberoftargets \gg \numberofherders$ uncontrolled agents (targets) in a periodic one-dimensional spatial domain $\mathcal{S} = [-\pi, \pi]$.

As done in \cite{di2025continuification}, we model each herder as a single integrator driven by the control input. Formally,
\begin{equation} \label{eq:herders}
    \dot{\singleherder}_i (t) = \controlaction_i (t), \quad \singleherder_{i}(0) = \singleherder_{i,0}, \quad i = 1, \dots, \numberofherders,
\end{equation}
where $\singleherder_i (t) \in \domain$ denotes the position of the $i$-th herder, and $\controlaction_i (t) \in [-\herdermaxvelocity, \herdermaxvelocity]$ is the corresponding control input.
We also define the vector containing the positions of all the herders as $\herders = \left[\singleherder_1, \dots, \singleherder_{\numberofherders} \right]$
and the vector containing their respective control actions as $\controlactions = \left[ \controlaction_1, \dots, \controlaction_{\numberofherders} \right].$
As in \cite{di2025continuification}, the targets are modeled as random walkers repelled from nearby herders. Formally, denoting the position of target $j$ as $T_j$, we have
\begin{equation} 
\begin{split}
    \de \singletarget_j (t)  = \frac{\interactionstrengthgain(t)}{\numberofherders + \numberoftargets} \sum_{i=1}^{\numberofherders} \interactionkernel \bigl(\{\singletarget_j (t), \singleherder_i (t)\}\bigr) \de t\\ + \sqrt{2 \diffusioncoefficient}\, \de B_j (t),
    \quad j = 1, \dots, \numberoftargets.
\end{split}
\label{eq:micro_target}
\end{equation}
The drift term models the interaction between herders and targets, occurring through a periodic interaction kernel $\interactionkernel(x):\domain\to\BB{R}$. We chose
\begin{equation}
\label{eq:interaction_kernel}
    f(x) = \frac{\mathrm{sgn}(x)}{\mathrm{e}^{2\pi/\kernelinteractionlength}-1} \left[\mathrm{e}^{\frac{2\pi-\vert x\vert }{\kernelinteractionlength}} - \mathrm{e}^{\frac{\vert x\vert}{\kernelinteractionlength}}\right],
\end{equation}
as done in \cite{maffettone2025leaderfollower}.
Here, $\{T_j, H_i\}$ denotes the wrapped distance between target $j$ and herder $i$ in the periodic domain.
As is common in shepherding applications \cite{long2020comprehensive, lama2024shepherding}, $\interactionkernel(x)$ is repulsive (i.e., $f(x)\mathrm{sign}(x)\geq0 \; \forall x \in \domain$), odd, and vanishes as the distance increases according to a characteristic interaction length $L>0$. 
Additionally, we assume that the strength of the repulsion can be modulated by a gain $\interactionstrengthgain(t)>0$. This is especially relevant where the herders are robotic swarms, such as in oil-spill containment \cite{zahugi2013oil}.
The diffusion term comprises independent standard Wiener processes ($B_j(t)$) with diffusion coefficient $\diffusioncoefficient>0$.

In the limit $\numberoftargets \to \infty$, mean-field theory~\cite{maffettone2025leaderfollower} allows us to describe the target population through its spatial density $\rho^T : \mathcal{S} \times \mathbb{R}^+ \to \mathbb{R}^+$, whose evolution follows the Fokker-Planck equation:
\begin{equation} \label{eq:fokker_planck}
    \targetdistribution_t (x,t) + \left[ \targetdistribution (x,t) \interactionstrengthgain(t) f(x)*\herderdistribution(x,t) \right]_x = D \targetdistribution_{xx} (x,t),
\end{equation}
where $*$ is the convolution operation, the subscripts $t$ and $x$ indicate the time and space partial differentiation, respectively. This equation is complemented with periodic boundary conditions, in compliance with $\domain$. 
Next, we define the empirical distribution of the herders,  $\herderdistribution : \domain \times \mathbb{R}^+ \to \mathbb{R}^+$, as
\begin{equation} \label{eq:herders_distribution_empirical}
    \herderdistribution (x,t) = \frac{\massofherders}{\numberofherders} \sum_{i=1}^{\numberofherders} \delta \bigl(x - \singleherder_i (t)\bigr).
\end{equation}
with $\massofherders$ being the total mass of herders. 
Given the definition of $\rho^T$, we can also define the total mass of targets as
\begin{equation}
    \massoftargets = \int_\domain \targetdistribution (x,t)\, \de x, \quad \forall t \in \timedomain.
\end{equation}
Note that $\massofherders + \massoftargets$ is constant for all $t \geq 0$, given our boundary conditions. Without loss of generality, we set  $\massofherders + \massoftargets = 1$.

We remark that, although $\herderdistribution(x,t)$ and $\targetdistribution(x,t)$ share the same domain, they are supported on different sets, namely $\mathrm{supp}(\herderdistribution(x,t)) = \{H_1(t), \dots, H_{\numberofherders}(t) \}$ and $\mathrm{supp}(\targetdistribution(x,t)) = \domain \subset \mathbb{R}$.
Unlike \cite{di2025continuification, maffettone2025leaderfollower}, where the herder population is large enough to be approximated by smooth densities governed by a Fokker–Planck equation, our sparse herder distribution remains a sum of Dirac measures. This leads to a coupled ODE–PDE system, where the individual herder dynamics (ODEs) drive the evolution of the target density (PDE) toward the desired configuration. %

The control problem can then be formulated as follows. 
\begin{problem} \label{prob:control_problem}
    Given the coupled ODE–PDE system described by Eqs.~\eqref{eq:herders} and \eqref{eq:fokker_planck}, let $\desiredtargetdistribution(x)$ be a desired target density solving the shepherding problem. 
    We wish to design the control inputs $\controlactions(t)$ for the herders' dynamics so as to indirectly steer the target density and achieve
    $\lim_{t \to \infty} \norm{\desiredtargetdistribution(x) - \targetdistribution(x,t)}_\Lnorm \leq \varepsilon$,
    within some tolerance $\varepsilon > 0$ due to the presence of sparse actuation.
\end{problem}
Here, $\norm{f(x)}_\Lnorm$ denotes the $\Lnorm$ norm of a function $f$ over the spatial domain $\domain$, defined as
$\norm{f(x)}_\Lnorm = \left( \int_{\domain} |f(x)|^2 \, \de x \right)^{1/2}.$
Since the goal of shepherding problems is to corral agents within a defined spatial region, we select the desired target distribution as a von~Mises distribution with concentration parameter $\kappa>0$, that is, 
\begin{equation} \label{eq:von_mises}
\desiredtargetdistribution (x) = \massoftargets \frac{\exp \bigl(\kappa \cos (x)\bigr)}{\int_\domain \exp \bigl(\kappa \cos (x)\bigr) \de x},
\end{equation}
which concentrates the targets around $x=0$. This choice is consistent with previous studies addressing shepherding control in the continuum setting \cite{maffettone2025leaderfollower, di2025continuification}.
Finally, we set $\numberofherders=2$ to investigate the case in which a very small number of herders is tasked to control a large target population.


\section{Control design}
To address the above Control Problem, we propose the control strategy illustrated in Figure~\ref{fig:control_arch}. The architecture is built around a \emph{macro–micro controller} trained via Proximal Policy Optimization (PPO) \cite{schulman2017proximal}. 
Recall that the herders’ positions influence the target distribution through the interaction kernel (see Eq.~\eqref{eq:fokker_planck}). The PPO controller computes the herder velocities $\controlactions(t)$ based on the desired target density $\desiredtargetdistribution(x)$, generating the velocity field $\velocityfield(x,t) := K(t) f(x) * \herderdistribution(x,t)$, which in turn shapes the evolving target distribution $\targetdistribution(x)$.
The controller therefore learns to steer the target distribution indirectly by optimizing the herders’ positions in closed loop, while treating the target density itself in open loop. This design reduces the infinite-dimensional PDE control problem to a tractable finite-dimensional Reinforcement Learning task, thereby eliminating the need for computationally expensive optimization.

The agent receives as input the positions of the herders in $\domain$. Specifically, we encode these positions as $\left[\cos(H_1), \sin(H_1), \dots, \cos(H_{\numberofherders}), \sin(H_{\numberofherders})\right] \in [-1,1]^{2\numberofherders}$. This choice is made to prevent chattering in the control inputs caused by numerical discontinuities at $x=\pm \pi$.
The Reinforcement Learning agent then outputs the herders velocities, given by the vector $\controlactions=\left[u_1, \dots, u_{\numberofherders}\right] \in [-\herdermaxvelocity, \herdermaxvelocity]^{\numberofherders}$. Note that this strategy is centralized, as the control inputs for all herders are decided by a single controller with global knowledge of the system.

The reward function driving the training could be chosen as:
\begin{equation}
\label{eq:reward_rhoTss_desired}
\hat r(t_k) = -k_1 \|\bar{\rho}^T(x) - \rho^{T}(x, t_k)\|^2_\Lnorm - k_2 \|\mathbf{u}(t_k)\|^2_2,
\end{equation}
where $t_k = k\Delta t$ is the discretized time-step, $\Delta t$ is the discretization step, $\norm{\cdot}_\Lnorm$ is the $\Lnorm$ functional norm over the domain $\domain$, $\norm{\cdot}_2$ is the Euclidean vector norm.
The first term penalizes the error between the desired and actual target distributions, while the second term regularizes the control effort to prevent excessive energy consumption and ensure smooth herder trajectories. The weights $k_1 \gg k_2$ prioritize accuracy over energy efficiency.

This approach would require computing the current target density at every time step, explicitly simulating the PDE governing the target dynamics and providing the high-dimensional density as input to the neural network.
However, this can be cumbersome due to the high-dimensionality of the network input layer.
To address this, we propose a more practical alternative based on a modified reward function of the form
\begin{equation}
\label{eq:reward_rhoTss}
r(t_k) = -k_1 \|\bar{\rho}^T(x) - \rho^{T,ss}(x; \mathbf{H}(t_k))\|^2_\Lnorm - k_2 \|\mathbf{u}(t_k)\|^2_2,
\end{equation}
where $\targetdistributionss(x; \mathbf{H}) : \domain \times \domain^{\numberofherders} \to \mathbb{R}^+$ is an estimation of the target density distribution at steady state parametrized at the herders position (see section \ref{sec:estimation_rhoss} for more details). 

\subsection{Steady-state target density estimation} \label{sec:estimation_rhoss}

At each time step $t_k$, the steady-state target density is estimated by fixing the herders’ positions, so that they generate a time-invariant velocity field acting on the targets. This estimate is then updated at the next time step based on the herders’ new positions.
Under this assumption, we compute $\targetdistributionss(x;\mathbf{H})$ by setting $\rho_t^\mathrm{T}(x) = 0$ in Equation~\eqref{eq:fokker_planck} and solving the resulting spatial ODE with periodic boundary conditions. This yields
\begin{equation} \label{eq:target_distribution_ss}
\targetdistributionss(x;\mathbf{H}) =
Z \exp \Biggl( \int_{-\pi}^x \frac{\interactionstrengthgain(t)\massofherders}{\diffusioncoefficient \numberofherders} \sum_{i=1}^{\numberofherders} \interactionkernel \bigl( \xi - \singleherder_i(t) \bigr) \mathrm{d}\xi \Biggr),
\end{equation}
where $Z$ is a normalization constant ensuring $\int_\domain \targetdistributionss(x;\mathbf{H}) \mathrm{d}x = \massoftargets$, for all $\mathbf{H} \in \domain^{\numberofherders}$.
Note that, due to the periodicity of the interaction kernel--and consequently of the time-invariant velocity field--the density $\targetdistribution(x,t)$ admits a unique and stable steady-state solution~\cite{gardiner2004handbook}, provided that the kernel is smooth. This condition can be easily satisfied by adopting a smooth approximation of~\eqref{eq:interaction_kernel}. 

\subsection{Interaction strength adaptation}
We complement the Reinforcement Learning agent with an interaction strength compensation to improve performance. We can exploit this additional degree of freedom to mitigate the limitations imposed by the small number of herders.
Specifically, recalling that the strength of the interaction between herders and targets in Equation \eqref{eq:fokker_planck} can be modulated through the gain $\interactionstrengthgain(t)$, we employ a heuristic gradient-descent compensation law 
\begin{equation} \label{eq:adaptive_K}
    \dot{\interactionstrengthgain}(t) = - \alpha \, \nabla_\interactionstrengthgain \norm{\targetdistributionss (x;\herders(t)) - \desiredtargetdistribution (x)}_\Lnorm,
\end{equation}
with initial gain $K(0)=1$.
Such compensation law allows the convergence of the interaction strength gain to a value that locally minimizes the target distribution mismatch.


\section{Numerical validation}
The system of herders and targets described in Equations \eqref{eq:herders} and \eqref{eq:fokker_planck} is simulated by using Forward Euler to integrate the ODEs and finite differences to integrate the PDE \cite{quarteroni2009numerical}.
The spatial domain is discretized uniformly with a step size $\Delta x$.
Two different time step sizes were used: a larger $\Delta t$ for the ODEs and a much smaller $\Delta t_\text{PDE}$ for the PDE, following the guidelines in \cite{quarteroni2009numerical} for numerical stability.
Table~\ref{tab:parameters} summarizes the parameters used for simulations.
The choice of parameters in Table~\ref{tab:parameters} reflects realistic constraints in robotic shepherding applications, with herder velocities bounded by physical actuator limits and diffusion coefficient chosen to model moderate environmental noise.

\begin{figure}[!t]
    \centering
    \includegraphics[width=0.7\linewidth]{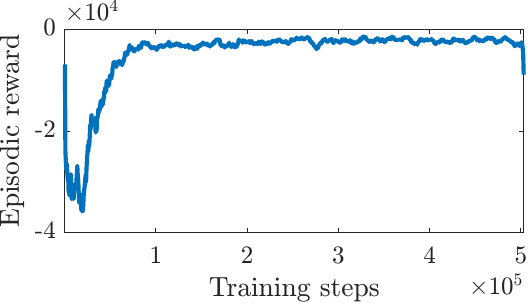}
    \caption{Episodic reward during the training process of the PPO agent. Values are smoothed with a moving average of width 20 steps.}
    \label{fig:training_ppo}
\end{figure}

\begin{table}[!t]
\centering
\begin{tabular}{lll}
\toprule
Symbol                      & Parameter                         & Value                     \\
\midrule
$\Delta x$                  & Spatial step size                 & $2\pi/250$   \\
$\Delta t$                  & Time step size for ODE simulation & $0.01$                    \\
$\Delta t_{\text{PDE}}$     & Time step size for PDE simulation & $0.0005$                  \\
$\timehorizon$              & Time horizon                      & 150                  \\
$\numberofherders$          & Number of herders                         & $2$           \\
$\diffusioncoefficient$     & Diffusion coefficient                     & $0.05$        \\
$\kernelinteractionlength$  & Kernel interaction length                 & $\pi$         \\
$\kappa$                    & Concentration of Von Mises distribution   & $16/\pi^2$  \\
$v_\mathrm{max}$      & Herders' maximum velocity                 & $3$           \\
$\massofherders$            & Mass of herders' population               & $0.3$         \\
$\massoftargets$            & Mass of targets population                & $0.7$         \\
$\alpha$                    & Adaptation law step size                  & $0.2$         \\
\bottomrule
\end{tabular}
\caption{Main parameters used to simulate the experiments. }
\label{tab:parameters}
\vspace{-0.8cm}
\end{table}

\begin{figure*}[!t]
    \centering
    \includegraphics[width=1\linewidth]{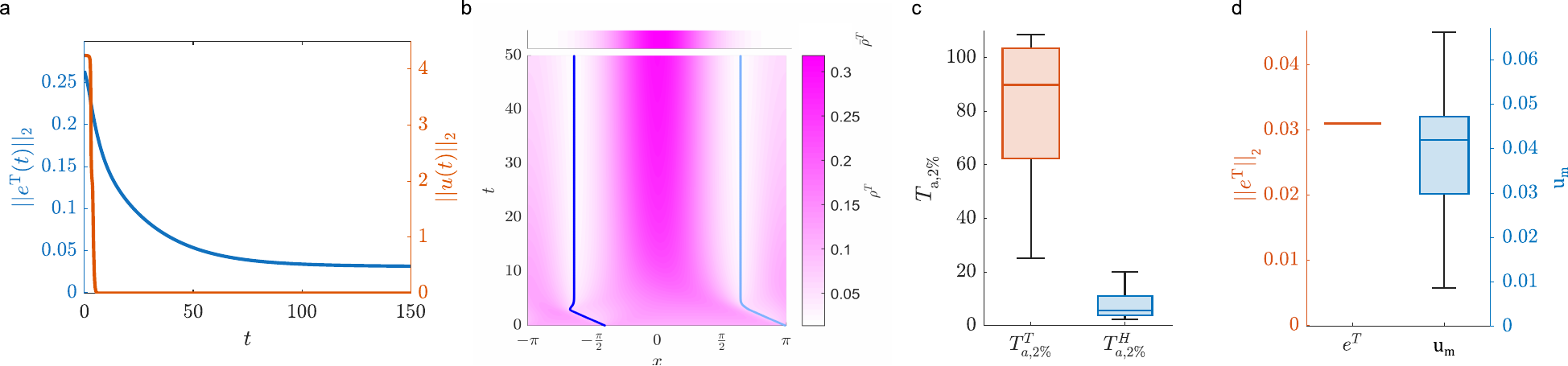}
    \caption{Performance of the controller using the reward function in~\eqref{eq:reward_rhoTss} and the compensation law in~\eqref{eq:adaptive_K}. 
    (a) $\mathcal{L}^2$ norm of the target distribution error $\targetdistributionerror$ (blue) and Euclidean vector norm of the control effort $\norm{\controlactions}_2$ (red) over time;
    (b) Top panel: Desired target density distribution $\desiredtargetdistribution$. Bottom panel: Evolution of $\targetdistribution$ and of the herders positions in space (x axis) and time (y axis) for a representative experiment. The shade of magenta is representative of the density of the targets. The dark and light blue lines describe the position of the herders at each time instant;
    (c) boxplots of the settling time (2\%) for the target population (red, left) and the herders (blue, right), calculated over 10 experiments with random initial conditions; and (d) boxplots of the steady-state target density error $\targetdistributionerror_{\mathrm{ss}}$ (red, left) and the control effort $\mathrm u_m$ (blue, right), calculated over 10 experiments with random initial conditions.  
    } 
    \vspace{-0.35cm}
    \label{fig:controller_performance}
\end{figure*}

PPO was implemented following the algorithm proposed in \cite{schulman2017proximal}. Both the actor and the critic are implemented as fully connected feed-forward neural networks.
Each network includes four hidden layers of 64 neurons with ReLU activations.
The actor network outputs the parameters of a Gaussian policy: the mean action, obtained through a hyperbolic tangent activation to ensure bounded values, and an independent learnable standard deviation.
During training, actions are sampled from this Gaussian distribution to encourage exploration, whereas during validation the deterministic mean action is applied directly.
At the beginning of each episode, the initial positions of the herders are sampled uniformly from the domain, i.e., $\singleherder_{i,0} \sim \C{U}(\domain)$, to ensure broad exploration.

The parameters of the reward function in Equation~\eqref{eq:reward_rhoTss} were set to $k_1 = 10$ and $k_2 = 0.01$, while the gradient descent gain in Equation~\eqref{eq:adaptive_K} was set to $\alpha = 0$ during training.
Figure~\ref{fig:training_ppo} illustrates the evolution of the episodic rewards during training, showing that they converge to a plateau corresponding to relatively high cumulative reward values, indicating effective learning and numerical stability.

Videos regarding the experiments and the hyperparameters used--refined starting from those in \cite{schulman2017proximal}--can be found at \hyperlink{https://github.com/SINCROgroup/Sparse-shepherding-control-of-large-scale-multi-agent-systems-via-Reinforcement-Learning}{https://tinyurl.com/ywhun59y}.

\subsection{Performance indices}
\label{sec:Metrics}
We define a set of metrics to assess the performance of the proposed control strategy. 
Specifically, we quantify the steady-state performance by computing the $\Lnorm$ function norm of the steady-state target density error, defined as
\begin{equation} \label{eq:metric_rho_error}
        e^{\mathrm{T,ss}} = \norm{\targetdistributionerror(x,\timehorizon)}_\Lnorm =  \norm{\desiredtargetdistribution (x) - \targetdistribution (x,\timehorizon)}_\Lnorm,
\end{equation}
where $\targetdistributionerror(x,t)$ represents the target distribution error and $\timehorizon$ is the final time of the experiment.

Additionally, we measure the overall control effort exerted by the herders by evaluating the average vector norm of $\controlactions(t)$, defined as
\begin{equation}
    \mathrm{u}_\mathrm{m} = \frac{1}{\timehorizon} \int_{0}^{\timehorizon} \norm{\controlactions (t)}_2 \de t,
\end{equation}
where $\norm{\cdot}_2$ here indicates the Euclidean vector norm.

Finally, we assess the transient performance using the $2\%$ settling time of herders ($\herderssettlingtime$) and targets ($\targetssettlingtime$) by setting:
\begin{equation}
    \begin{split}
       & \herderssettlingtime = \max_{i=1,\dots,\numberofherders}\inf_{t_i} \{ 0 \le t_i \le \timehorizon :\\ 
        & \abs{\singleherder_i (\tau_i) - \singleherder_i(\timehorizon)} \leq 0.02 \abs{\singleherder_i(0) - \singleherder_i(\timehorizon)}, \, \forall \tau_i \geq t_i\},    
    \end{split} 
\end{equation}

\begin{equation}
    \begin{split}
        \targetssettlingtime &= \inf_{t} \{ t > 0: \abs{\,\norm{\targetdistributionerror(x,\tau)}_\Lnorm - e^\mathrm{T,ss}\,} \le \\
        & \le 0.02 \abs{\,\norm{\targetdistributionerror(x,0)}_\Lnorm - e^\mathrm{T,ss} \,}, \, \forall \tau \geq t \}
    \end{split}
\end{equation}

\subsection{Controller performance and comparison}
To assess the performance of the trained control agent, we conducted ten numerical experiments, each with $\singleherder_{i,0}\sim \mathcal U(\domain)$. 
The gain of the gradient descent in Equation~\eqref{eq:adaptive_K} is set to $\alpha = 0.2$ and the gradient is computed numerically.

Figures~\ref{fig:controller_performance}a–b illustrate a representative experiment. In particular, in Figure~\ref{fig:controller_performance}a, the $\mathcal{L}^2$ norm of the density error settles at $0.031$ after approximately $90$ time units, while the control input norm rapidly decreases to zero within about $4.23$ time units. 
Figure~\ref{fig:controller_performance}b shows the evolution of the target density during a single experiment over space (x-axis) and time (y-axis). The density is represented by shades of magenta, while the dark and light blue lines indicate the herders’ positions. 
We can observe that the herders quickly reach a steady-state configuration, after which the target density converges to a distribution qualitatively similar to the desired one $\desiredtargetdistribution(x)$.

Overall, this example demonstrates effective reward choice, successfully abating both terms in Equation~\eqref{eq:reward_rhoTss}.

We further evaluate the controller’s performance using the indices defined in Section \ref{sec:Metrics}. As shown in Figure \ref{fig:controller_performance}c, the herders exhibit a significantly shorter settling time than the targets, indicating that they rapidly converge to a steady-state configuration. This enables them to generate an effective velocity field acting on the targets. Moreover, Figure \ref{fig:controller_performance}d shows that the residual error of the target density consistently stabilizes around $0.031$, with a very low standard deviation. This demonstrates that the Reinforcement Learning agent has learned a consistent strategy for optimally positioning the herders to produce the most effective velocity field on the targets, regardless of the herders’ initial conditions. Finally, the average control effort remains approximately $0.042$ across all experiments, which is remarkably low compared to $\herdermaxvelocity$.

\begin{figure}[!t]
\centering
\includegraphics[width=0.7\linewidth]{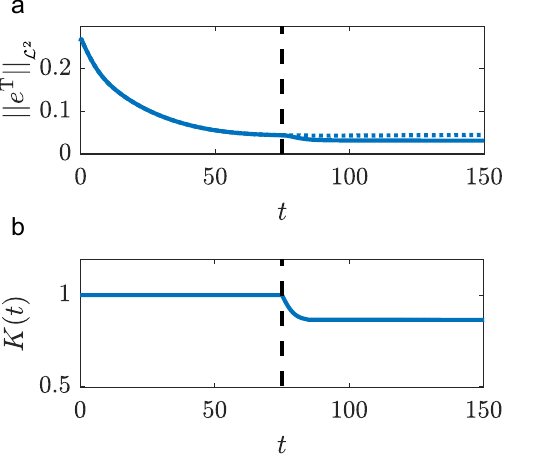}
\caption{Evolution of the $\norm{\targetdistributionerror(x,t)}_\Lnorm$ when activating the compensation law for gain $\interactionstrengthgain (t)$ for $t \ge 75$ (indicated as a vertical black dotted line). (a) $\mathcal{L}_2$ norm of $\targetdistributionerror$ with (solid line) and without (dotted line) compensation, highlighting the advantages of the compensation law. (b) time evolution of the interaction strength gain $\interactionstrengthgain (t)$.}
\label{fig:adaptive_K}
\vspace{-0.35cm}
\end{figure}

To further assess the impact of the compensation law introduced in \eqref{eq:adaptive_K}, we isolate its effect by fixing $K=1$ for the first $75$ time units and then activating the adaptive law for the remainder of the simulation. The results in Figure \ref{fig:adaptive_K} show that the controller trained with PPO reaches a steady state after approximately $58.97$ time units, with a residual error of $e^{\mathrm{T,ss}} = 0.045$. Once the compensation law is activated, $\interactionstrengthgain(t)$ adapts dynamically, leading to a further reduction in the steady-state error and achieving an approximate $31\%$ decrease in the distribution error norm $\norm{\targetdistributionerror}_\Lnorm$ within $7.47$ time units.

We wish to emphasize that we have omitted explicit comparisons between our architecture and existing methods for solving sparse indirect control problems (e.g., \cite{albi2016invisible, elamvazhuthi2021controllability, albi2025micromacro}), as our work relies on different modeling assumptions.

\subsection{Robustness evaluation}

We numerically evaluated the robustness of the control algorithm against external disturbances. Specifically, we tested the controller’s robustness against constant additive disturbances in the herders’ dynamics and additive measurement noise. The former assesses the controller’s sensitivity to bounded unmodeled dynamics, while the latter accounts for imperfections in sensor measurements.

\begin{figure*}[!t]
    \centering
    \includegraphics[width=1\linewidth]{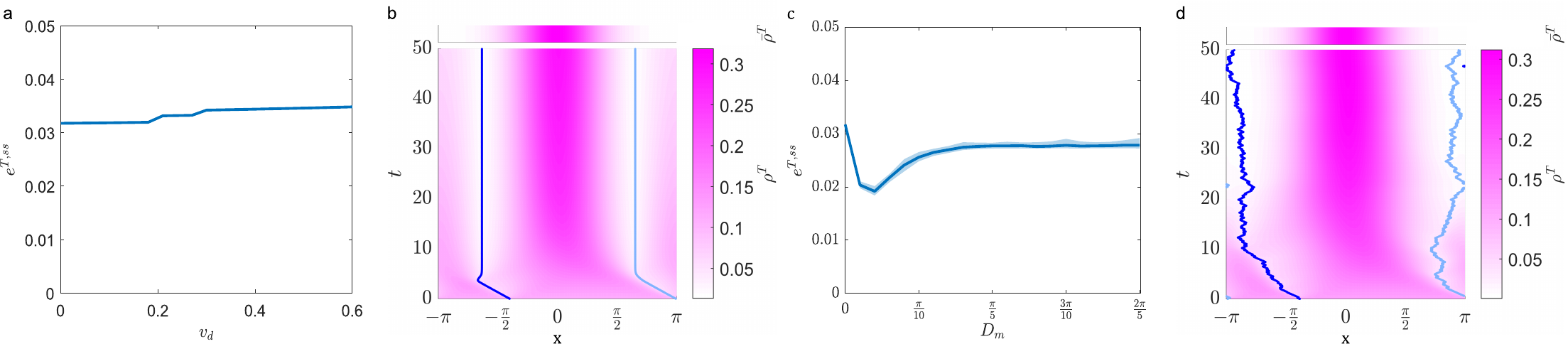}
    \caption{%
    Robustness analysis of the proposed control strategy against constant disturbances (panels a, b) and measurement noise (panels c, d). 
    (a) Steady-state error $\targetdistributionerror$$^{,ss}$ varying the constant disturbance on the herders dynamics. 
    (b) Top panel: Desired target density distribution $\desiredtargetdistribution$ in space. Bottom panel: Evolution of the targets density and of the herders positions in space (x axis) and time (y axis) for a representative experiment given $v_d=0.6$. The shade of magenta is representative of the density of the targets. The dark and light blue lines describe the position of the herders at each time instant. 
    (c) Mean (solid blue line), 10th and 90th percentiles (shaded blue) of the steady state target error varying the amplitude of the measurement noise over 100 numerical experiments. 
    (d) Top panel: Desired target density distribution $\desiredtargetdistribution$ in space. Bottom panel: Evolution of the targets density and of the herders positions in space (x axis) and time (y axis) for a representative example given $D_m=2\pi/5$. The shade of magenta is representative of the density of the targets. The dark and light blue lines describe the position of the herders at each time instant.}
    \vspace{-0.35cm}
    \label{fig:controller_robustness}
\end{figure*}

Figure \ref{fig:controller_robustness}a shows the steady-state target density error, denoted by $e^{\mathrm{T,ss}}$, as a function of a constant additive disturbance $v_d \in [0, 0.6]$ applied to the herder dynamics \eqref{eq:herders}. The maximum disturbance amplitude is empirically set to $v_d = 0.6$, corresponding to $20\%$ of the maximum herder velocity $\herdermaxvelocity$. The disturbance is discretized in $1\%$ increments, resulting in $21$ sample points.
The numerical experiments indicate a maximum increase of $9\%$ in the steady-state error $e^{\mathrm{T,ss}}$ when $v_d = 0.6$, demonstrating the controller’s robustness to this class of disturbances.
The effect of $v_d = 0.6$ can also be observed in Figure \ref{fig:controller_robustness}b, which illustrates the spatial and temporal evolution of the target density, and still closely matches the desired one at steady-state. This demonstrates that, even in the worst-case scenario considered, the target density remains largely unaffected by the disturbance introduced.
In particular, the error norm converges to $e^{\mathrm{T,ss}} = 0.034$, which is comparable to the steady-state error obtained in the absence of any disturbance.

Next, we evaluated the controller’s robustness to additive measurement noise by introducing white Gaussian noise with a standard deviation $D_m \in [0, 2\pi/5]$ and performing $100$ experiments for each value of $D_m$. This range was selected so that the measurement noise had a standard deviation of up to $20\%$ of the total domain size, and $D_m$ was discretized in $1\%$ increments. 
Interestingly, the performance of the measurement noise slightly improves the controller's performance. 
This effect occurs because the noise introduces small fluctuations in the herders’ positions, leading to a smoother target density profile that more closely matches $\desiredtargetdistribution$.
When $D_m  = 2\pi/5$, which corresponds to the worst-case scenario considered, the target error norm converges on average to $e^{\mathrm{T,ss}} = 0.03$, which is comparable to the error obtained in the absence of measurement noise. These results demonstrate that the steady-state target density is not significantly affected by measurement noise, confirming that the controller successfully achieves its objective despite perturbations.
Figure~\ref{fig:controller_robustness}d illustrates the herders’ positions and the target density in a numerical experiment with $D_m = 2\pi/5$. In this scenario, despite the strong randomness in the measured position, the herders are still able to steer the target density towards a density close to $\desiredtargetdistribution$.

\subsection{Computational effort}
The training of the Reinforcement Learning agent required ${\sim\!27}$ minutes on a Ryzen 7 3700x CPU with 32GB of RAM, without GPU acceleration.
Training was performed on Windows 10 with Python 3.12.3.
Inference requires $4.1\times 10^{-4}$ seconds per step using the same hardware.

\section{Conclusions}
We formulated a sparse indirect control problem and proposed an optimal learning-based solution. Using PPO, we trained a controller to minimize the $\mathcal{L}^2$ error between the desired and expected steady-state density profiles while keeping the control effort low. By modulating the interaction strength $\interactionstrengthgain(t)$, we further reduced the residual steady-state error and partially compensated for the limited number of herders.
Our results demonstrate that the proposed learning-based approach effectively controls large-scale target populations while remaining robust to constant disturbances and measurement noise. 

Future work will focus on extending the framework to incorporate real-time target density feedback, potentially through distributed sensing~\cite{lorenzo2025decentralizeda} or learned density encoders.
Another direction is the transition from centralized to decentralized architectures using multi-agent Reinforcement Learning~\cite{gronauer2022multi}. This paradigm shift would improve scalability by allowing the same policy to be shared among herders, thus avoiding the need to retrain the neural network whenever the number of herders changes. Moreover, we also expect higher robustness to individual herder failures and enabling adaptive collective decision-making.
Furthermore, establishing theoretical guarantees for the convergence and stability of the learned policies remains an important open challenge.
Another relevant direction concerns extending the proposed strategy to higher spatial dimensions, enabling the framework to address more realistic spatial environments for robotic swarms. Specifically, experimental validation on physical robot swarms will be crucial to assess the practical feasibility of the approach and identify limitations not captured in simulation.


\bibliographystyle{IEEEtran}
\bibliography{IEEEabrv,references}

\end{document}